\documentclass[12pt]{article}
\usepackage{psfrag}
\usepackage{graphicx}
\usepackage{rotating}
\usepackage{longtable}
\usepackage{supertabular}
\usepackage{latexsym}
\usepackage{amssymb}
\usepackage{float}

\title{Role of topological defects in the phase transition of modified XY model
: A Monte Carlo study\\[8mm]
}
\author{Suman Sinha
\footnote{E-mail: ssinha@research.jdvu.ac.in}
 ~and
Soumen Kumar Roy
\footnote{Corresponding author. E-mail: skroy@phys.jdvu.ac.in,
Tel: +91 9331910161; fax: +91 33 24146584}\\
Department of Physics,\\
Jadavpur University, Kolkata - 700032, INDIA}
\date{  }

\begin{document}

\maketitle
\begin{abstract}
Monte Carlo simulation has been performed on a classical two dimensional XY-model with a modified 
form of interaction potential to investigate the role of topological defects on the phase transition
exhibited by the model. In simulations in a restricted ensemble without defects, the system appears
to remain ordered at all temperatures. Suppression of topological defects on the square plaquettes
in the modified XY-model leads to complete elimination of the phase transition observed in this model.
\end{abstract}
\\
\section{\bf Introduction}
\begin{figure}[tbh]
\begin{center}
\resizebox{120mm}{!}{\rotatebox{-90}{\includegraphics[scale=1.2]{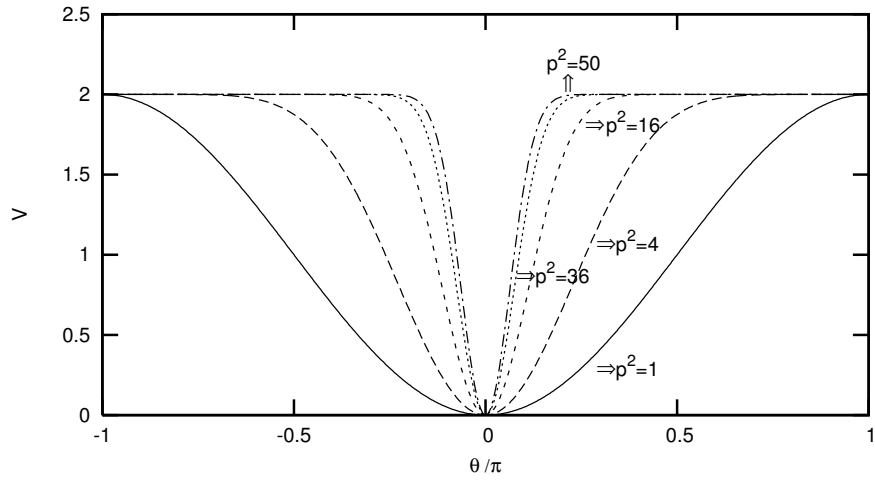}}}
\end{center}
\caption{The potential function of Eqn. (1) is shown for different values of $p^2$.}
\label{pot}
\end{figure}

In 1984, Domany, Schick and Swendsen \cite{ds} introduced an extension of the two-dimensional (2D) XY-model
where the classical spins (of unit length), located at the sites of a square lattice and free to rotate
in a plane, say the XY plane (having no z-component) interact with nearest-neighbors through a 
modified potential
\begin{equation}
V(\theta_{ij})=2\Big[1-\Big(\tt cos^2{\frac{\theta_{ij}}{2}}\Big)^{p^2}\Big]
\label{eqn1}
\end{equation}
where $\theta_{ij}$ is the angle between the nearest neighbor spins and $p^2$ is a parameter used to
change the shape of the potential. For $p^2=1$, the potential reduces to that of a conventional XY-
model while with the increase in $p^2$, the potential well gets narrower with a width $\sim \pi /p$ and
for $\theta \gtrsim \pi /p$ it is essentially constant at $V(\pi)=2$. The shape of the potential is 
shown in Fig.~\ref{pot} for several values of $p^2$. The conventional 2D XY-model does not possess
any true long range order \cite{merwag} and it is known that in this model, non-singular
spin wave excitations alone cannot destroy the quasi long range order (QLRO). However, the presence of
topological defects leads to a QLRO-disorder phase transition, now familiar as the Kosterlitz-Thouless 
(KT) phase transition. Kosterlitz and Thouless \cite{kt,kos}predicted that topological singularities 
are both necessary and sufficient for the QLRO-disorder phase transition in the 2D XY-model and using 
a renormalization group (RG) approach they established that the phase transition is mediated by 
unbinding of vortices and anti-vortices which are stable topological defects in this system. 
The phase with QLRO is characterized by a slow algebraic decay of the spin-spin
correlation function whereas a fast exponential decay is observed in a disordered system. The KT 
transition in the 2D XY model was unambiguously confirmed numerically by A. C. Irving and R. Kenna
\cite{irk}.

The modified XY-model of Eqn. (\ref{eqn1}) has been analyzed by a number of investigators 
\cite{ds,him,garel,min1,min2,min3,jmn,es,rhs,ssskr1,ota} and they all were of the opinion that it exhibits a
first order phase transition for large values of $p^2$. However, some investigators \cite{knops,him1,mila}
attempted to interpret the Monte Carlo (MC) results for the first order phase transition in other ways. In 
a recent work \cite{ssskr2} we have shown by performing extensive numerical simulations on relatively 
larger lattice sizes (upto $192 \times 192$) that the modified XY-model for large values of $p^2$ 
exhibits first order phase transition and all the finite size scaling rules for a first order phase 
transition were seen to be obeyed accurately. van Enter and Shlosman \cite{es,es2} provided a rigorous  
proof of a first order phase transition in various SO(n)-invariant n-vector models which have
a deep and narrow potential well and the model under investigation is a member of this general
class of systems.

It was argued in Ref.~\cite{ds} that more than one type of excitations (i.e., topological excitations as
well as vacancy excitations) may play a role in changing the nature of the phase transition, as one would 
become relatively more important than the other by the alteration of the potential. Later Himbergen argued 
that topological excitations alone are sufficient to account for both the continuous and the first order 
phase transitions, but in a qualitatively different manner \cite{him}. Algebraic topology or homotopy theory 
in the study of defects \cite{mer,kenna} has a wide application in the physics of phase transition. 
In the present paper, we investigate the role of topological defects in the phase
transition of the modified XY-model under consideration. We are specifically interested in inquiring
whether the first order phase transition in the modified XY-model is defect driven or not. In other
words, if in the absence of the role played by topological defects, would one observe the same 
order-disorder phase transition as the one found in the system with topological defects? If suppression
of the defects changes the nature of the phase transition or eliminates it altogether, one may conclude
that topological defects are necessary to describe the phase transition correctly.

The present work was motivated to a great extent by the work of Lau and Dasgupta \cite{ld} and Dutta and
Roy \cite{sdskr}. Lau and Dasgupta have shown that hedgehogs (point singularities in the 3D Heisenberg model)
are necessary for the phase transition in 3D Heisenberg model. They observed that if the formation of 
topological defects is suppressed in the 3D Heisenberg model, the system remains ordered at all temperatures
and the transition to the disordered phase disappears altogether. Dutta and Roy performed MC simulations
on planer $P_2$ and $P_4$ models which are known to exhibit a continuous and a first order phase transition
respectively. It was noticed that the suppression of the defects in these models leads to a total
disappearance of the phase transitions observed in these systems \cite{sdskr}. 
Other work, along the same line, that should be
mentioned in this context, is that of Lammert $et. al$ \cite{lam}. These authors have shown in a MC study
that the nature of the nematic-isotropic transition changes when one suppresses the formation of the 
stable line defects, called the disclination lines.

In the present work, we have found that topological defects are necessary for the existence and also
for proper description of phase transitions exhibited by these class of models. We also find that the 
change in the nature of the phase transition that is observed with the change in the value of $p^2$
is due to a change in the role played by the topological defects in these systems.

We arrange the rest of the paper as follows : Section 2 describes the simulational procedures used in 
the present work. The results and discussions are then presented in Section 3 followed by the concluding
remarks.

\section{\bf The simulation details}
\label{sd}
In order to study the behavior of the topological excitations and the role of topological defects in the 
phase transitions exhibited by the model under investigation, we have used the conventional Metropolis single
spin update algorithm \cite {metro,nb} with some modifications in our MC simulation. We have found that while
simulating a continuous lattice spin model using standard Metropolis algorithm, we need to adjust a parameter
very carefully to generate a new a configuration. This parameter determines the amplitude of the random angular
displacements of the spins and the results become very sensitive to the value of this parameter. In order to 
get rid of this difficulty of choosing the parameter, we generate a new spin configuration following the 
prescription of Wolff \cite{wlf}. We take a random unit vector $\vec r$ and a spin flip $\vec s \rightarrow
\vec s^{~\prime}$ is defined as 
$\vec s^{~\prime}=\vec s - 2\left(\vec s, \vec r\right)\vec r$ where $(\vec s, \vec r)$ is the dot product of
$\vec s$ and $\vec r$. Apart from this method of generating a new configuration, the rest of the algorithm is
the standard Metropolis algorithm. Defining a spin flip in that way, our modified Metropolis algorithm is free
from tuning any adjustable parameter while simulating a lattice spin model with continuous energy spectrum
while the conditions of ergodicity and detail balance remain fulfilled. The 
Metropolis algorithm runs as follows: first we choose a spin at random from within the specified range. Then the
change in energy $\Delta E$ associated with an attempted move is calculated. If $\Delta E\leq0$, the attempted
move is accepted. If $\Delta E>0$, the attempted move is accepted with probability $ {\tt exp} (-\Delta E/T)$
where $T$ is the dimensionless temperature. 

The average defect pair density is calculated in the following way. A vortex (anti-vortex) is a topological
defect in which the angle variable $\theta$, specifying the direction of the order parameter, changes by
$2\pi (-2\pi)$ in one circuit of any closed contour enclosing the excitation core. In order to trace out 
the topological defects, we consider a square plaquette in the physical space. Let $\vec s_1$, 
$\vec s_2$, $\vec s_3$ and $\vec s_4$
 be the four spins at the corners of the square plaquette. The angles between these adjacent spins are
calculated with proper sign and these are then summed algebraically to find the total angle. The square
plaquette is said to enclose a vortex (topological charge $Q=1$) when the sum equals $2\pi$ or more
precisely very close to $2\pi$, taking into account the possible numerical errors. The square plaquette is
said to enclose a anti-vortex (topological charge $Q=-1$) if the sum equals $-2\pi$. If the sum is zero, there
is no topological defect in the plaquette. Average defect pair density (taking into consideration both vortices
and anti-vortices) is calculated as the thermodynamic average of the absolute value of the vorticity summed over
the entire lattice divided by the total number of spins. In this method, it is ensured that the net 
topological charge is always equal to zero in a system with periodic boundary conditions. It should be mentioned
here that the smallest part of the system in real space that enclose a $Q=\pm 1$ point defect is a triangle. We
can divide each elementary square plaquette diagonally into two triangles. One could thus consider a triangular
plaquette in the physical space as well to trace out topological defects. We have tested that the total number
of topological charges in the entire lattice remains the same whether we choose a square plaquette or a triangular
plaquette. Only topological charges of strength $Q=\pm 1$ are considered since they are energetically favorable.
In our investigation of the equilibrium behavior of topological defects near phase transition, we carried out
simulations on system sizes with linear dimension $L=16$, $32$, $48$ and $64$ with periodic boundary conditions.
$10^5$ Monte Carlo sweeps (MCS) were used for equilibration and $10^6$ MCS were used for calculating 
thermodynamic averages. One MC sweep is said to be completed when the number of attempted single spin moves
equals the total number of spins in the system. The values of $p^2$ taken to study the variation of average 
defect pair density with $p^2$ are $4$, $9$, $16$, $25$, $36$, $50$, $64$, $81$ and $100$.

In order to implement the procedure of the suppression of topological defects in our model, a 
``chemical potential'' term associated with the topological charges is included \cite {ld,sdskr}. 
The modified Hamiltonian in the simulation is given by
\begin{equation}
H_m=\sum_{{\langle}ij{\rangle}}2\Big[1-\Big(\tt cos^2{\frac{\theta_{ij}}{2}}\Big)^{p^2}\Big]
+ \lambda \sum_{ijkl} |\it Q_{ijkl}|
\label{eqn2}
\end{equation}
where $\theta_{ij}$ is the angle between the nearest neighbor spins $i$, $j$ and $|Q_{ijkl}|$ is the absolute
value of the charge enclosed by the square plaquette. A positive value of $\lambda$ ensures that the formation 
of the charges becomes expensive in terms of energy. So for positive $\lambda$, this ``chemical potential'' 
term has the effect of suppressing configurations containing defects. In simulation with this modified 
Hamiltonian, a calculation of $\Delta E$, the energy change associated with an attempted move of a spin, 
involves calculations of the changes in topological charges associated with the four unit square plaquettes which
share the spin under consideration. For almost complete suppression of the defects, the value of $\lambda$
was chosen to be between $5$ and $20$ irrespective of temperature. The $\lambda \rightarrow \infty$ limit
of Eqn. (\ref{eqn2}) indicates an ensemble in which configurations containing topological defects are not 
allowed. We started our simulation in that restricted ensemble with a configuration in which all the spins 
are aligned parallel to one another, ie, there is no topological defect. The restricted simulations were 
carried out by using the modified Metropolis spin update algorithm described earlier in this section. 
We performed our restricted simulations on system sizes with linear dimension $L=16$, $32$, $64$ and $96$ 
with periodic boundary conditions. The $\lambda =0$ corresponds to an unrestricted simulation, where no 
suppression of topological defects take place.

For the purpose of calculating various thermodynamic quantities, we have used multiple histogram reweighting
technique of Ferrenberg and Swendsen \cite{fs}. In the restricted simulations, $10^6$ MCS were taken for
equilibration and $10^7$ MCS were used for computing the raw histograms (both energy and order parameter
histograms). The value of $p^2$ was taken to be $50$ in order to carry out the restricted simulations.

\section{\bf Results and discussions}
In this section we present in detail the results obtained from our simulations.
\subsection{\bf Behavior of topological excitations near phase transition}
\begin{figure}[tbh]
\begin{center}
\resizebox{120mm}{!}{\rotatebox{-90}{\includegraphics[scale=1.2]{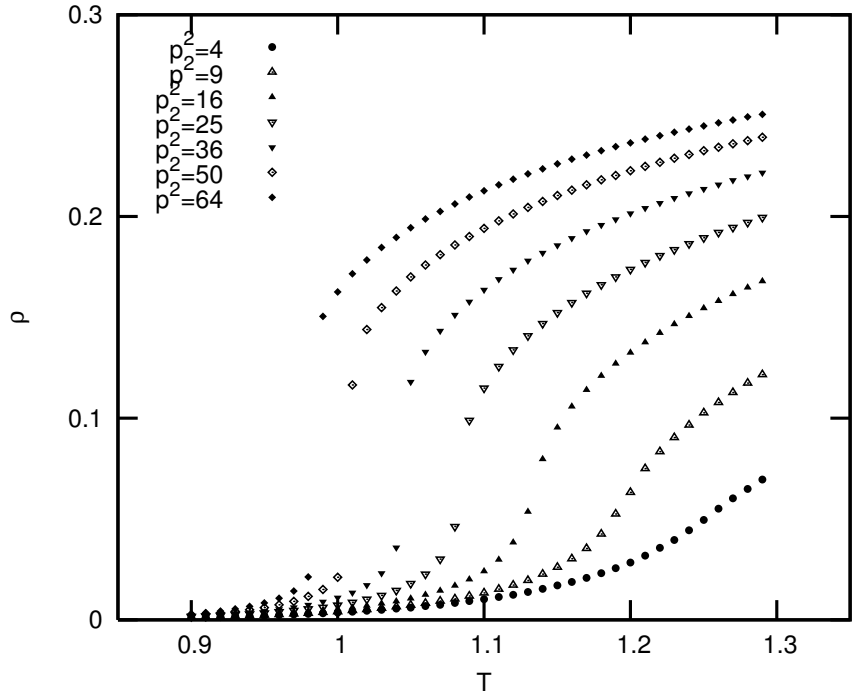}}}
\end{center}
\caption{Average defect pair density $\rho$ plotted against dimensionless temperature $T$
for $L=64$ for various values of $p^2$}
\label{defdenvst}
\end{figure}

We used the method described in Sec.~\ref{sd} to determine the average defect pair density ($\rho$) of 
the system. The variation of the $\rho$ with the dimensionless temperature $T$ is shown in
Fig.~\ref{defdenvst} for several values of the parameter $p^2$. The average defect pair density is found 
to increase sharply as $T$ increases through the transition temperature $T_c(p^2)$ and appears to exhibit 
a sharp jump at $T_c(p^2)$, particularly for $p^2 \gg 1$. Fig.~\ref{defdenvst} indicates that $T_c(p^2)$ 
decreases as the values of $p^2$ increases. It is evident from Fig.~\ref{defdenvst} that for larger values 
of $p^2$, at some temperature $T_c(p^2)$, vortices suddenly appear in great numbers and a first order phase 
transition takes place which is in accordance with the explanation of Himbergen \cite{him}. It may be noted 
that Jonsson, Minnhagen and Nylen  \cite{jmn} also performed MC simulations on a 2D XY-model with a modified 
potential, which is essentially equivalent to that of Eqn. (\ref{eqn1}), and interpreted the first order 
transition to be of vortex unbinding type. 

\begin{figure}[tbh]
\begin{center}
\resizebox{120mm}{!}{\rotatebox{-90}{\includegraphics[scale=1.2]{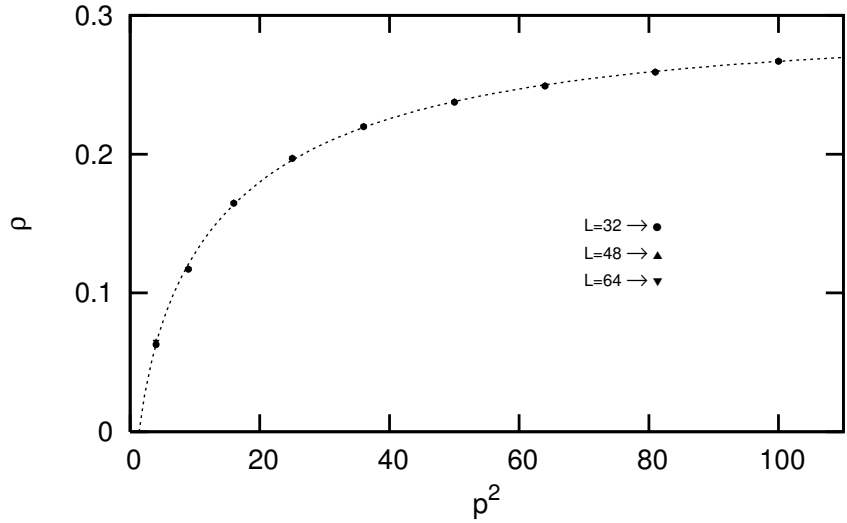}}}
\end{center}
\caption{Average defect pair density $\rho$ plotted as a function of $p^2$ at $T=1.28$ for three system sizes.
 The best fit corresponds to $L=64$. The errorbars are smaller than the dimension of the symbols used for
plotting.}
\label{p2vsdd}
\end{figure}

We have also studied the behavior of topological excitations with the parameter $p^2$. The average defect pair
density ($\rho$) as a function of the parameter $p^2$ is plotted in Fig.~\ref{p2vsdd} for three different 
system sizes at a temperature $T=1.12$ which is above the transition temperature of the model for $p^2=50$. 
We observe that above the transition temperature, the data for $\rho$ versus $p^2$ are nicely fitted by 
the following expression
\begin{equation}
\rho(T)=\rho_{\tt max}-\alpha(T){\tt exp}(-\gamma \sqrt p^2)
\label{eqn3}
\end{equation}
Eqn. (\ref{eqn3}) takes into account both vortices and anti-vortices. The values of $\rho_{\tt max}$ for 
the three system sizes are listed in Table 1. There is no significant system size dependence of the 
parameters and it may be noted that $\rho$ increases with $p^2$. In the limit $p^2 \rightarrow \infty$, the
system contains only vortex excitations. This means that in the high $p^2$ limit, the system must be disordered
even at very small temperatures and consequently the transition temperature must be very low. This is the reason
behind the decrease in the transition temperature with increase in $p^2$. In this context, we refer to the work  
 of Romano $et. al$ \cite{romano} who, in a MC study of 2D generalized XY-model with 
spin component $n=3$, discussed the nature of the phase transition with the variation of a generalized parameter.

\begin{figure}[tbh]
\begin{center}
\resizebox{120mm}{!}{\rotatebox{-90}{\includegraphics[scale=1.2]{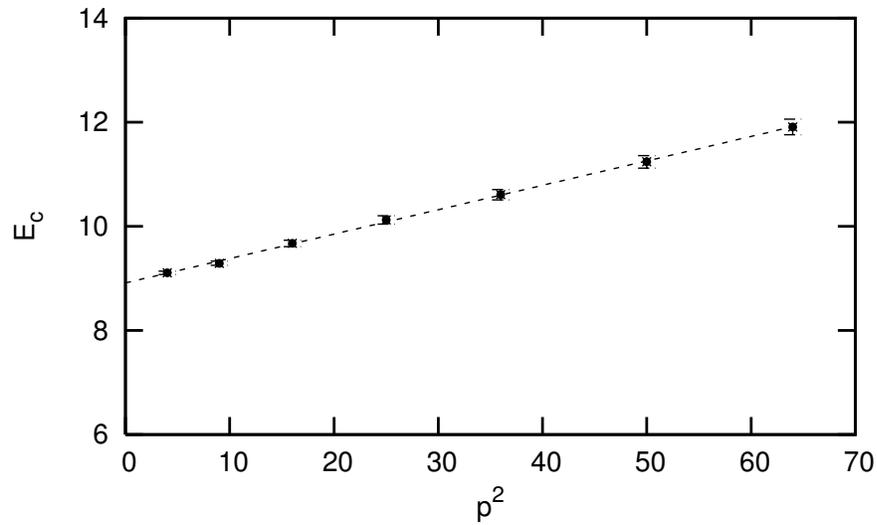}}}
\end{center}
\caption{The defect core energy $E_c$ plotted against the parameter $p^2$ with the linear fit
represented by the dotted line. The error bars are of the dimension of the symbols used for plotting.}
\label{p2vscorengy}
\end{figure}

We have also calculated the defect core energy ($E_c$) for various values of $p^2$. Because there is always 
a positive energy cost $E_c$ associated
with the creation of a vortex core, thermally excited vortices in thermal equilibrium always contribute
terms proportional to ${\tt exp} (-E_c/T)$ to the partition function. Therefore the total number of 
topological charges ($n$) shows an exponential behavior ${\tt exp} (-E_c/T)$ at low temperatures and is 
given by
\begin{equation}
n=n_0 {\tt exp}(-E_c/T)
\label{eqn4}
\end{equation}
Taking natural logarithm on both sides of Eqn. (\ref{eqn4})
\begin{equation}
\ln n=\ln n_0 - \frac{E_c}{T}
\label{eqn5}
\end{equation}
The defect core energy $E_c$ for each $p^2$ is determined from the linear fit of the plot $\ln n$ versus
$1/T$ and Fig.~\ref{p2vscorengy} shows the plot of variation of $E_c$ versus $p^2$. 
The defect core energy $E_c$ for $p^2=50$ model (which is known to exhibit a strongly first order
phase transition) is found to be $11.911 \pm 0.15$ while that for $p^2=1$ model (which is known to exhibit
a continuous phase transition) is found to be $7.560 \pm 0.019$. This is in apparent contradiction with 
the main finding of Saito \cite {saito} who, in a MC study of a system of interacting dislocation vectors,
 predicted a continuous phase transition due to large core energy and a first order phase transition due 
to small core energy.

\subsection{Restricted simulations with no defects}
Before presenting our results of restricted simulations, we briefly define the thermodynamic quantities
that we have evaluated. The MC simulations were carried out with the modified Hamiltonian given by
Eqn. (\ref{eqn2}) where the new term acts as a ``chemical potential'' for the defects.

The specific heat $C_v$ is evaluated from the energy fluctuations
\begin{equation}
C_v=\frac{1}{N}{\frac{\displaystyle\left({\langle}H^2{\rangle}-{\langle}H{\rangle}^2\right)}{\displaystyle T^2}}
\label{eqn6}
\end{equation}
where $T$ is the dimensionless temperature and $N=L^2$ is the total number of spins.
The average order parameter is given by
\begin{equation}
{\langle}P_1{\rangle}={\langle}\tt cos \phi {\rangle}
\label{eqn7}
\end{equation}
where $\phi$ is the angle that a spin makes with the preferred direction of orientation and the average
is over the entire sample.
The first rank pair correlation function is defined as
\begin{equation}
G_1(r)={{\langle}\left(\tt cos \theta_{ij}\right){\rangle}}_r
\label{eqn8}
\end{equation}
where $i$ and $j$ are two spins separated by a distance $r$.
The second rank pair correlation function is defined as
\begin{equation}
G_2(r)={{\langle}P_2\left(\tt cos \theta_{ij}\right){\rangle}}_r
\label{eqn9}
\end{equation}

We did not find any evidence for a phase transition from the ordered to the disordered phase at any 
temperature in the simulations of the restricted ensemble where configurations containing defects are
not allowed.

\begin{figure}[tbh]
\begin{center}
\resizebox{120mm}{!}{\rotatebox{-90}{\includegraphics[scale=1.2]{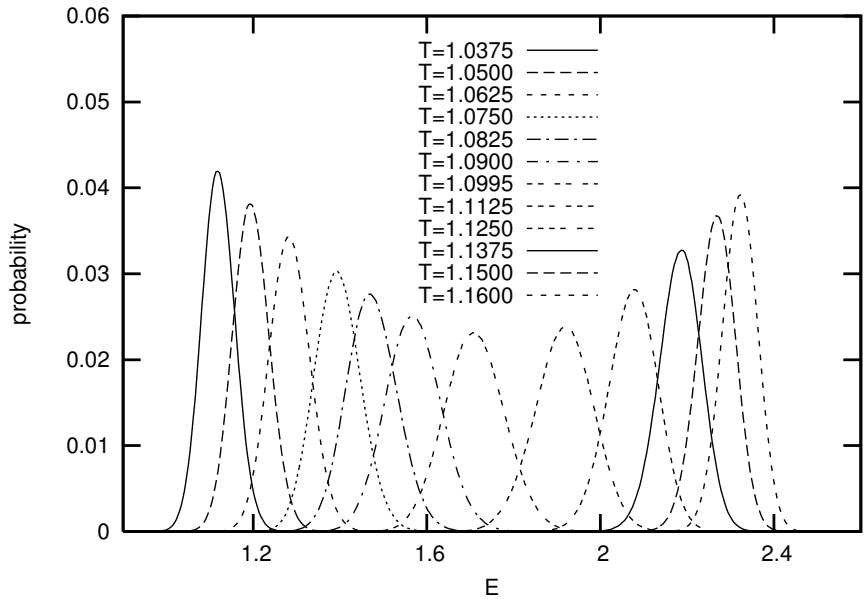}}}
\end{center}
\caption{The histograms for $E$, the average energy per particle generated for the $64 \times 64$
lattice for the $p^2=50$ model at the 12 temperatures indicated.}
\label{enghist}
\end{figure}

The energy histograms obtained for $L=64$ are shown in Fig.~\ref {enghist}. For this lattice, simulations 
were performed at 12 different temperatures ranging from $T=1.0375$ to $T=1.1600$. It is evident from the 
energy histograms that the dual peak nature of the histograms as obtained for an unrestricted simulation 
\cite{ssskr2} disappears in a restricted simulation. It is known that the dual peak nature of the histograms 
is a signature of a first order transition where two phases can coexist at a given temperature.

\begin{figure}[tbh]
\begin{center}
\resizebox{120mm}{!}{\rotatebox{-90}{\includegraphics[scale=1.2]{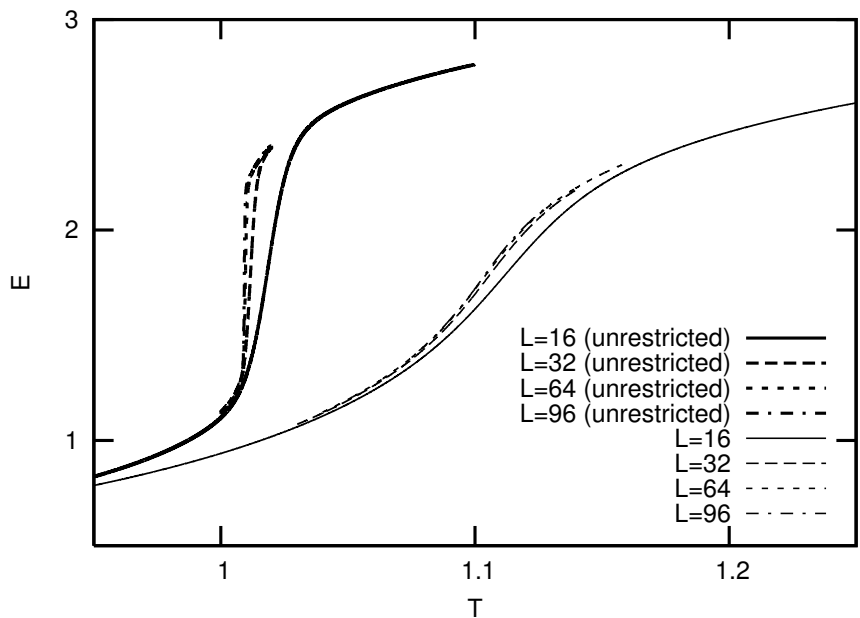}}}
\end{center}
\caption{The average energy per particle $E$ plotted against dimensionless temperature $T$ for different
lattice sizes. The thick curves (on the left) correspond to unrestricted simulations.}
\label{eng_compare}
\end{figure}

\begin{figure}[tbh]
\begin{center}
\resizebox{120mm}{!}{\rotatebox{-90}{\includegraphics[scale=1.2]{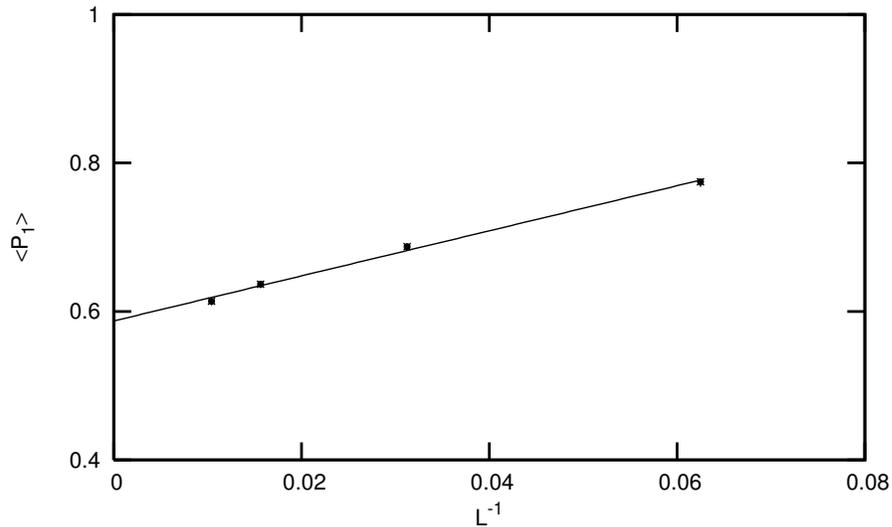}}}
\end{center}
\caption{The plot of order parameter ${\langle}P_1{\rangle}$ at $T=1.12$ vs $1/L$. The straight line is the best linear fit
to the data. The errorbars are smaller than the dimension of the symbols used for plotting.}
\label{opvsl}
\end{figure}

Fig.~\ref{eng_compare} shows the temperature dependence of the average energy (E) for a number of lattices, 
as obtained by applying the histogram reweighting technique. For comparison, the same plots for unrestricted 
simulations (where no defect is suppressed) are shown by thick lines in the same figure. It is evident from 
the figure that the energy changes only gradually and smoothly with temperatures for the restricted simulations 
while a sharp variation of the same with temperature is observed in the unrestricted case. The average value 
of the order parameter ${\langle}P_1{\rangle}$ defined in Eqn. (\ref{eqn7}), is, always nonzero for a finite 
size system. Hence we have studied the system size dependence of ${\langle}P_1{\rangle}$. The values of 
${\langle}P_1{\rangle}$ at $T \rightarrow \infty$ versus $1/L$ are plotted in Fig.~\ref{opvsl} and the system 
size dependence is, in fact, well fitted by the form ${\langle}P_1{\rangle}=P_0+a/L$ with $P_0=0.587 \pm 0.005$ 
and $a=3.035 \pm 0.141$. It is clear from Fig.~\ref{opvsl} that there is no indication of 
${\langle}P_1{\rangle}$ extrapolating to zero in the thermodynamic limit $L \rightarrow \infty$ and thereby 
suggesting a state with long range ferromagnetic order.

\begin{figure}[tbh]
\begin{center}
\resizebox{120mm}{!}{\rotatebox{-90}{\includegraphics[scale=1.2]{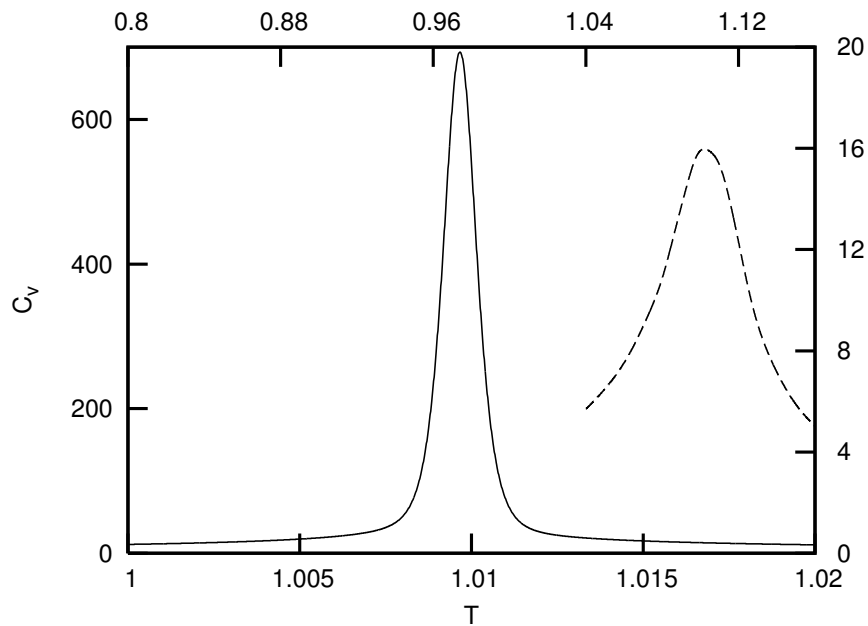}}}
\end{center}
\caption{The specific heat $C_v$ plotted against dimensionless temperature $T$ for restricted ensemble 
(represented by dashed line) for $L=64$. The same is plotted for unrestricted ensemble (shown by solid line).
For clarity the graphs are plotted in two different scales. The bottom X-axis and left Y-axis are chosen
to plot the results for unrestricted simulations while the top X-axis and right Y-axis are chosen to plot
the same for restricted simulations.}
\label{cv_compare}
\end{figure}

\begin{figure}[tbh]
\begin{center}
\resizebox{120mm}{!}{\rotatebox{-90}{\includegraphics[scale=1.2]{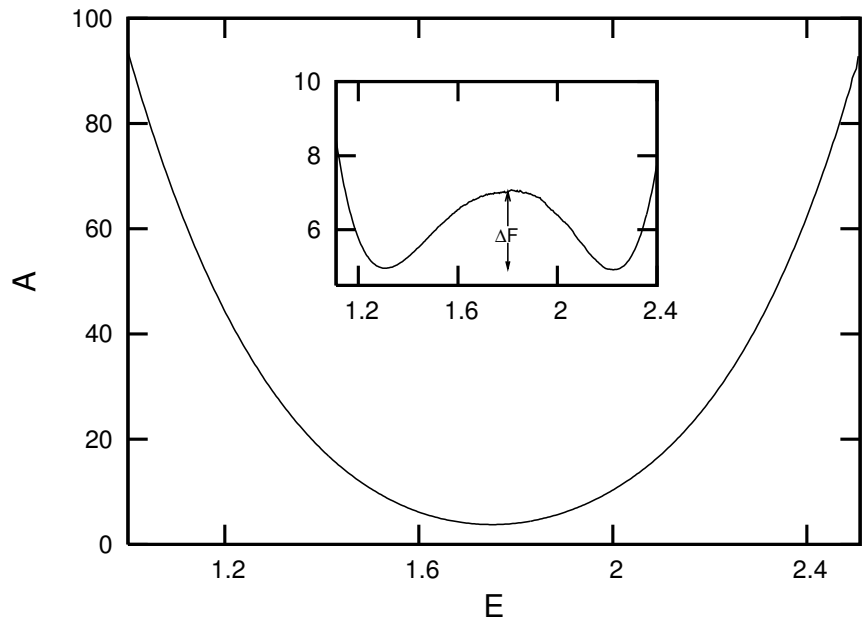}}}
\end{center}
\caption{The free energy $A$ generated by the multiple histogram reweighting technique plotted against
energy per particle for $L=64$ in the restricted ensemble. Absence of a double well structure is to be noted. 
The inset shows the same plot for an unrestricted ensemble where the presence of a double well structure is
observed.}
\label{prob_compare}
\end{figure}

The specific heat $C_v$ was obtained from the energy fluctuation relation (Eqn. (\ref{eqn6})). 
 The specific heat data for $L=64$ in a restricted ensemble are shown by dashed 
line in Fig.~\ref{cv_compare} where the results for the unrestricted case are also shown by solid line for 
comparison. For clarity the data for the restricted and the unrestricted simulations are plotted in two 
different scales. While $C_v$ has a large peak height $(\sim 700)$ at the transition temperature in the 
unrestricted case, which presumably is a signal of a phase transition in a finite system, in the defect-free 
case the peak height $(\sim 16)$ is drastically reduced and almost disappears in comparison with the normal 
case (where no defect is suppressed). We would like to argue that, in the restricted ensemble, the existence 
of a peak in $C_v$ of insignificant height (compared to that of an unrestricted ensemble) over the temperature 
range cannot be a sign of a phase transition. These may be attributed to the fact that complete suppression of 
topological defects is never possible, there always exists a small number of residual charges in the system. 
We have also calculated the free energy like quantity $A$ from the energy histograms. It is defined as 
$A(E;\beta,L,\mathcal N)=-\ln N(E;\beta,L)$ where $N(E;\beta,L)$ is the histogram count of the energy 
distribution. Fig.~\ref{prob_compare} shows the plot of the quantity $A$ against $E$ for $L=64$. The inset of 
Fig.~\ref{prob_compare} shows the same plot for the original model (Eqn. (\ref{eqn1})) where a double well 
structure of equal depth at the transition temperature signals a first order transition. We observe the absence 
of any such double well structure in $A$ when defects are suppressed. We would be inclined to conclude from 
the results of $C_v$ and $A$ that the defect free phase exhibits no phase transition at all.

\begin{figure}[tbh]
\begin{center}
\resizebox{120mm}{!}{\rotatebox{-90}{\includegraphics[scale=1.2]{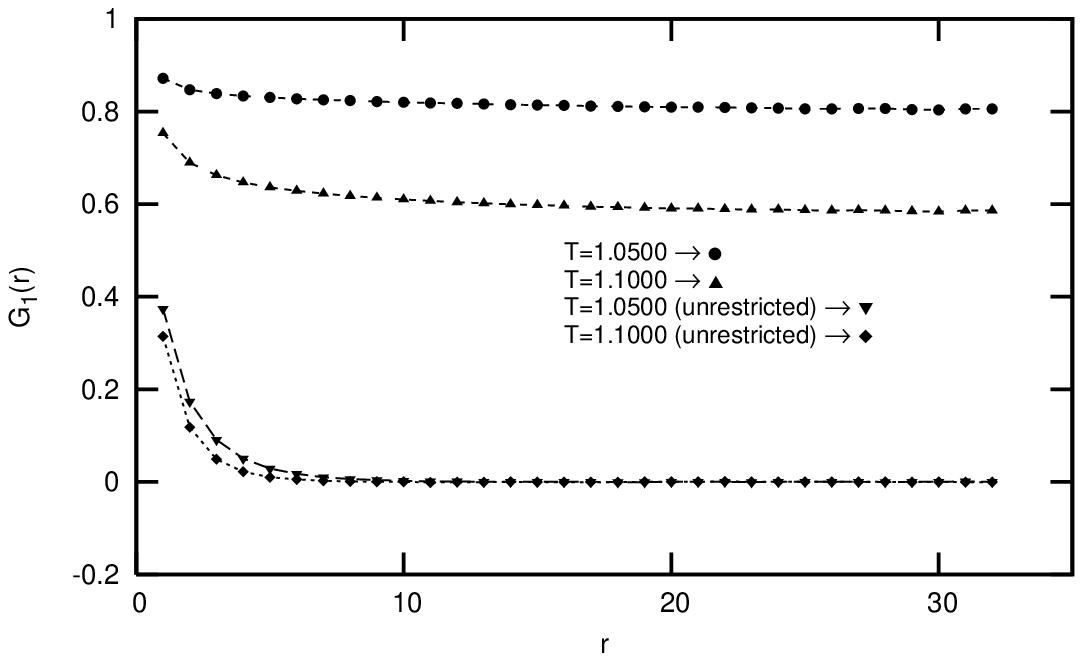}}}
\end{center}
\caption{The plots of the pair correlation function $G_1(r)$ against $r$ for the $64 \times 64$ lattice for
the temperatures indicated. The curves are plotted for $r$ ranging upto $L/2$.}
\label{angcor1}
\end{figure}

\begin{figure}[tbh]
\begin{center}
\resizebox{120mm}{!}{\rotatebox{-90}{\includegraphics[scale=1.2]{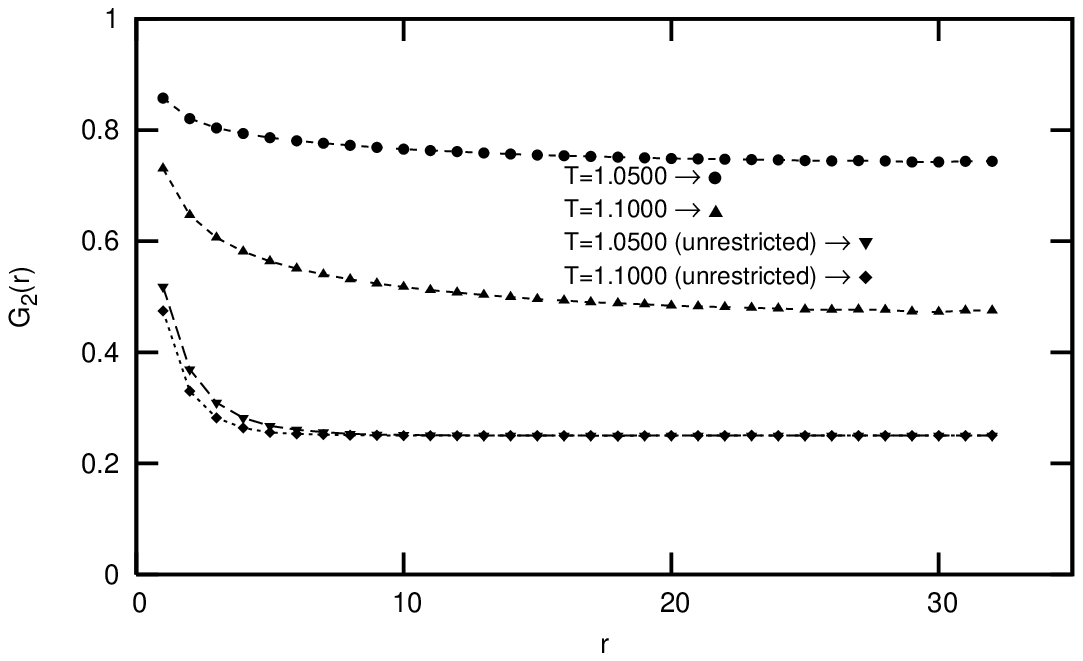}}}
\end{center}
\caption{The plots of the pair correlation function $G_2(r)$ against $r$ for the $L=64$ lattice for
the temperatures indicated. The curves are plotted for $r$ ranging upto $L/2$.}
\label{angcor2}
\end{figure}

We now turn to pair correlation functions defined earlier in this section. Fig.~\ref{angcor1} shows the plot of 
$G_1(r)$ against $r$ for $L=64$ in the restricted as well as the unrestricted cases. The first rank pair
correlation function $G_1(r)$ for $p^2=50$ at temperatures $T=1.0500$ and $T=1.1000$, which is much higher than 
the transition temperature of the original model, decays exponentially to zero in the unrestricted
simulations, as it should, in the complete absence of long range order and a best fit with
$G_1(r)=\alpha {\tt exp} (-\delta r)$ yields $\alpha=0.745\pm 0.01$ and $\delta=0.700 \pm 0.01$ for 
$T=1.0500$ and $\alpha=0.806\pm 0.009$ and $\delta=0.944 \pm 0.009$ for $T=1.1000$. For the simulation
where the defects are suppressed, $G_1(r)$ decays algebraically and a best fit with $G_1(r)=ar^{-b}+f$ yields
the parameter $a=0.064 \pm 0.003$, $b=0.175 \pm 0.016$ and $f=0.806 \pm 0.001$ for $T=1.0500$ and 
$a=0.019 \pm 0.009$, $b=0.271 \pm 0.018$ and $f=0.590 \pm 0.001$ for $T=1.1000$. It may be noted that the
parameter $f$ is the asymptotic value of the pair correlation function. The next higher order correlation
function $G_2(r)$ against $r$ for $L=64$ is plotted in Fig.~\ref{angcor2} at $T=1.0500$ and $T=1.0100$ for 
both the cases. The results indicate that $G_2(r)$ decays algebraically in both the cases and long range 
order prevails in the system via higher order correlation functions.

\begin{figure}[tbh]
\begin{center}
\resizebox{120mm}{!}{\rotatebox{-90}{\includegraphics[scale=1.2]{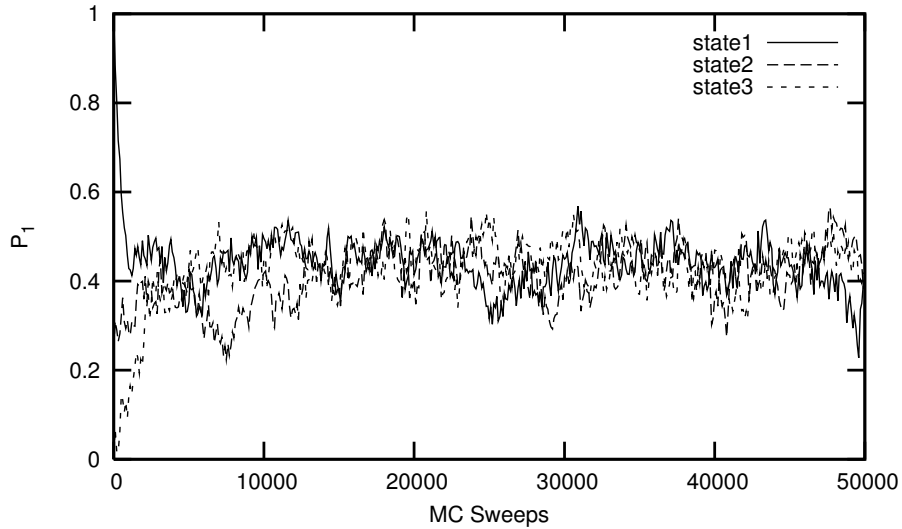}}}
\end{center}
\caption{The evolution of the order parameter at $T=1.1800$ for $L=64$ lattice after suppressing the defects using
$\lambda=20$ for three different initial configurations: $P_1=0.999$, $P_1=0.244$ and $P_1=0.012$. The final
values of the order parameter have the same average value.}
\label{opvsmc}
\end{figure}

We now need to address the question of phase space connectivity before arriving at the conclusion that
topological defects are indeed necessary for the phase transition. Since we have used large values of
$\lambda$ in our restricted simulations in order to suppress the evolution of topological defects, we
have to demonstrate that the observed behavior is not caused by trapping of the system 
in a small region of phase space with nonzero ${\langle}P_1{\rangle}$. Any MC study is guaranteed to 
generate appropriate ensemble averages if there is a path connecting any two points in the phase space
with nonzero transition probability. We have investigated the phase space connectivity by observing
the evolution of the order parameter and energy with MC sweeps. The connectedness is satisfied if the
observed quantities for different initial states converge to the same final value. In Fig.~\ref{opvsmc}, we have
shown that for $L=64$, after suppressing the defects (by using $\lambda=20$) on the square plaquettes,
the final values of the order parameter is same for three different initial configurations. 
 This observation ensures that we can use a value of $\lambda$ upto $20$
without violating the phase space connectivity and the observed non-vanishing of ${\langle}P_1{\rangle}$
is not a result of trapping of the system in the phase space.

\section{\bf Conclusions}
It is established in this paper that topological defects play a very crucial role in the phase transitions
exhibited by the models we discussed. We have observed that the average defect pair density grows rapidly
with the increase in $p^2$ (which increases the non-linearity of the potential well). For high $p^2$, 
the potential well becomes narrower so that there is an insufficient increase in the defect density
at low temperatures and then, at a certain temperature, they suddenly appear in the system in great
numbers. Therefore it may be thought that for larger values of $p^2$ the class of models, we have investigated, 
behave like a dense defect system and give rise to first order phase transition as has been predicted by 
Minnhagen \cite{min1,min2,min3}. It has also been observed that the first order transition is eliminated totally
when configurations containing topological defects are not allowed to occur and the system appears 
to remain ordered at all temperatures. Hence topological defects are necessary to account for the first 
order phase transition for larger values of $p^2$.

Another point which must be mentioned before ending this section is the performance of the Metropolis 
algorithm with the modification discussed earlier in this paper. The modification makes us free from 
tuning any adjustable parameter while simulating a continuous model and this has resulted in the model
being successfully simulated.

\section{\bf Acknowledgment}

        The authors acknowledge the award of a CSIR (India) research grant 03(1071)/06/EMR-II. 
One of us (SS) acknowledges the award of a fellowship from the same project. SS also thankfully
acknowledges his colleague Subhrajit Dutta for helpful discussions.

\begin{table}[!ht]
\caption{parameters for the fit of $\rho(T)=\rho_{\tt max}-\alpha(T){\tt exp}(-\gamma \sqrt p^2)$ for different $L$}
\begin{center}\
\begin{tabular}{|c|c|c|c|}
\hline
$L$ &$\rho_{\tt max}$ &$\alpha(T)$ &$\gamma$ \\
\hline
$32$ &$0.2866\pm 0.002$ &$0.411\pm 0.005$ &$0.301\pm 0.008$ \\
\hline
$48$ &$0.2876\pm 0.002$ &$0.406\pm 0.006$ &$0.297\pm 0.009$ \\
\hline
$64$ &$0.2878\pm 0.002$ &$0.406\pm 0.006$ &$0.296\pm 0.009$ \\
\hline
\end{tabular}
\end{center}
\end{table}

\end{document}